\documentclass[showpacs,preprint,amsmath,amssymb,prb]{revtex4}

\usepackage[dvips]{graphicx}   
\usepackage{dcolumn}           
\usepackage{bm}                

\newcommand{\figurewidth}{0.33\textwidth}

\begin{document}
\title{Strong Asymmetric Effect of Lattice Mismatch on Epilayer Structure in Metal Thin Film Deposition}
\author{Pai-Yi Hsiao}   \email[Corresponding author. E-mail: ]{pyhsiao@ess.nthu.edu.tw} 
\author{Zhuo-Han Tsai}
\author{Jia-Hong Huang} \email[E-mail: ]{jhhuang@ess.nthu.edu.tw}
\author{Ge-Ping Yu}     \email[E-mail: ]{gpyu@ess.nthu.edu.tw} 

\affiliation{Department of
Engineering and System Science, National Tsing Hua University,
Hsinchu, Taiwan 300, R.O.C.}

\date{\today}

\begin{abstract} We investigate the hetero-epitaxial growth of thin film
deposited on a (001) substrate via molecular dynamics simulations, using six
fcc transition metals as our modeling systems.  By studying the radial
distribution function in the film layers, we demonstrate the importance of the
sign of lattice mismatch on the layer structure.  For positive lattice
mismatches, the film favors pseudomorphic growth, whereas for negative
mismatches, a sharp transition happens within the first few monolayers of adatoms
and the film layers are transformed into the close-packed (111) structure.
We propose a way to quantify the compositional percentage of different planar
structures in an epilayer, and demonstrate the strong asymmetric effect between
the tensile and compressive cases of deposition.  How temperature affects the
asymmetry is also discussed.  \end{abstract}


\maketitle

How to epitaxially grow thin films on top of a substrate of different
crystalline structure is a challenging task and continually attracts great
attention of researchers due to its wide applications~\cite{epitaxy04, ohring}.
It is commonly acknowledged that to grow a high-quality single-crystal film in
a system where the lattice constant of the grown material and the substrate
differs is very difficult.  Therefore, a small lattice mismatch is generally
recognized as a determinative factor for epitaxial growth, together with other
factors such as surface/interface free energy, chemical bondings, temperature,
\textit{etc.}~\cite{ohring}.  In hetero-epitaxial growth, an elastic strain is
built up due to mismatched structure.  Misfit dislocation is thus formed
in the film layers to release the elastic energy while the elastic strain
approaches a certain degree, which allows us to define a critical thickness of
epitaxy.  In the classical continuum
theory~\cite{frank49,vandermerwe63,matthews75}, the elastic energy is simply
determined by the square of the lattice mismatch defined by $f=(a_{s}-a_{f})/a_{f}$
where $a_{f}$ and $a_{s}$ are the unstrained lattice constants of the film and the
substrate, respectively.  As a consequence, the critical thickness depend
only on $|f|$. This is why the epitaxial growth of thin film was 
usually examined regardless of the sign of $f$~\cite{matthews75b,stringfellow82,nix89}.
Nevertheless, scientists have shown that the sign of mismatch does have a
significant effect on the structure and the formation of misfit dislocations,
demonstrated by experiments and numerical studies~\cite{maree87,lu05,zhou06}.
This effect can be attributed to the tensile-compressive asymmetry, originated
from the anharmonicity of the atomic interaction~\cite{trushin02&03}.
Furthermore, surface energy also plays an important role in the film growth and
roughening.  It is known that the (111) surface has the lowest free energy for
face-centered cubic (fcc) materials.
Therefore, an epitaxial growth on the (100) surface for fcc metals will compete
with the formation of the (111) structure.  Different kinds of film structure
growing on a (100) substrate have been investigated in
literatures~\cite{bocquet97,rizzi03,mitlin04}.  Nonetheless, a systematic study
of the positive and negative lattice mismatch on the structure of a film layer
has rarely been conducted.  The information is missed: to what degree can the
tensile-compressive anharmonicity break the symmetry of the epitaxial growth
predicted by the continuum theory?  To answer this question, we chose six
transition metals of fcc crystalline structure as our modeling systems and studied
epitaxial growth of one metal on the other by means of molecular dynamics (MD)
simulations.  By using the radial distribution function, we were able to
calculate the percentage of different surface structure in a film layer and
studied the effect of lattice mismatch quantitatively. 

The six transition metals that we chose are the elements Ni(3.52\AA),
Pd(3.89\AA), Pt(3.92\AA), in the group 8B, and Cu(3.61\AA), Ag(4.09\AA),
Au(4.08\AA), in the group 1B.  The corresponding lattice
constant~\cite{kittel05} has been given, following the element symbol.  We
studied the thin film deposition of a metal F onto a metal S where F and S were
chosen from the six elements.  There are totally 36 combinations, which give
the value of lattice mismatch $f$ ranging from $-14\%$ to $16\%$.  We studied
the film growth on the (001) plane of a substrate and denoted the system by the
notation ``F/S(001)''.  The simulations were performed using a modified MD
package, LAMMPS~\cite{note-lammps}, with a time step equal to 0.001 picosecond
(ps).  The interaction between atoms is modeled by the embedded-atom
method~\cite{foiles86}.  The size of the substrate is of 10 unit cells in the
[100] ($x$-) and [010] ($y$-) directions; thus, 200 atoms constitutes an atomic
layer of the substrate.  There are 9 atomic layers in a substrate.  The atoms
in the two bottom layers are fixed on the lattice positions.  Above them, six
layers are thermal layers subject to a temperature control by velocity
rescaling method.  The top layer of the substrate is free of the temperature
control.  Four temperatures were studied: 80K, 300K, 600K and 900K.  In order to
simulate an infinite surface, periodic boundary condition was applied in $x$-
and $y$-directions.

In simulating the depositional growth, the adatoms were randomly dropped, one
by one, on the top of the substrate surface, 100{\AA} height above which, with
an initial downward velocity corresponding to an incident energy 0.1 eV.  This
incident energy stays in the range of a typical deposition by evaporation.  
The deposition rate is one atom per 5 ps. Since one monolayer (ML) contains 
$N_{\rm ML}=200$ atoms in our study, the growth rate is 1 ML per nanosecond,
which is several orders of magnitude higher than in molecular beam epitaxy. 
This choice is due to the limitation of standard MD techniques, 
restricted by today's computing power.
However, the high deposition rate used here does not invalidate the present study. 
In laser deposition techniques, such as those used in Ref.~\cite{lu05}, the 
fluxes are relatively high, approaching to our setup.  
To clarify the effect of high deposition rate, we have measured the relaxation time  
for an adatom after collision with the surface. 
We found that it is less than 1 ps, consistent with theoretical 
predictions~\cite{mccarroll63}. Therefore, the atom has enough time to relax to, 
at least, a local free energy minimum before coming in the next atom by this deposition rate.  
Another problem with high deposition rate is that there is no time for 
activation of many interlayer processes in film growth, such as
exchange diffusion and multiple-atom concerted diffusion~\cite{kellogg90}.
To take into account these infrequent complex processes, more sophisticated method such as 
hyperdynamics or kinetic Monte Carlo simulations should be applied~\cite{voter02}, 
where the flux can be set to a few orders of magnitude smaller than in standard MD 
simulations. Nevertheless, it requires much more computing resources and efforts. 
For a systematic study of the asymmetric effect in heteroepitaxial growth, 
we neglected these infrequent processes as the first-order approximation and 
approached this problem using standard MD simulations, in order to capture 
the fundamental picture of such systems. The simulation results will be benchmarked 
later with experimental data obtained by other groups to verify the validness of 
this setup. We deposited 16 monolayers (ML) of adatoms for each studied case.
In order to efficiently remove the heat brought in by the adatoms, the thermal layers 
were extended upward, one layer by one layer, each time when one ML of adatoms has 
been deposited onto the surface. 

\begin{figure}[htbp]
\includegraphics[width=\figurewidth,angle=270]{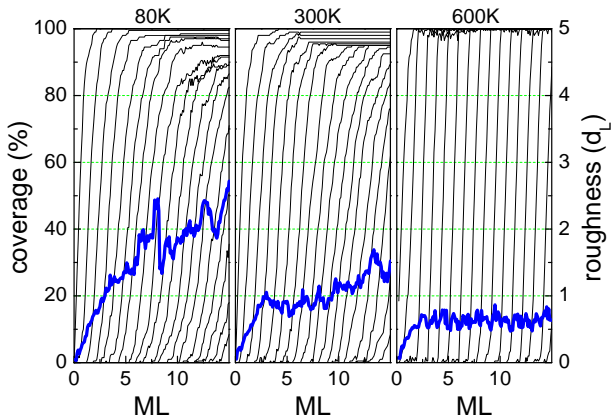}
\caption{$C_{\ell}(m)$ and $R(m)$ for Cu/Ni at  80K, 300K and 600K.  The thin
curves appearing from left to right at a given temperature denote, in turn, the
coverage for $\ell=1$, $2$, $\dots$ epilayer.  The values are read from
the left $y$-axis.  $R(m)$ is plotted in thick curve.  The values are read
from the right $y$-axis. } \label{coverage&roughness} 
\end{figure}

We first studied the Cu/Ni system. The mismatch is  $-2.49\%$.  The negative
mismatch means that the film suffers from a compressive strain.  We monitored
the coverage of each epilayer during deposition by
$C_{\ell}(m)=N_{\ell}(m)/N_{\rm ML}$ where $N_{\ell}(m)$ is the number of
adatoms in the $\ell$-th epilayer while $m$ ML of adatom has been deposited.
The results are shown in Fig.~\ref{coverage&roughness} at three 
temperatures: 80K, 300K and 600K.

$C_{\ell}(m)$ is zero at beginning.  It rapidly increases, one layer after
another, and saturates to some value.  For the first few layers, this value is
nearly $100\%$, which indicates a good coverage.  However, as $\ell$ increases,
$C_{\ell}(m)$ saturates to a value significantly small then 1, suggesting the
formation of voids in the high epilayers.  We found that the higher the
temperature, the less the voids will be formed.  It can be attributed to the
high kinetic energy of the adatoms due to the high substrate temperature.  With
high kinetic energy, the adatoms can easily overcome energy barrier and
migrate to the voids, flattening the film with good coverage.  At 600K, the
layer coverage is nearly $100\%$ and no void is formed in the epilayers.
Moreover, only a single $C_{\ell}(m)$ curve increases at any given moment at
this temperature. The film thus grows layer by layer; it is the Frank-van der
Merwe growth mode.  On the other hand, at 80K and 300K, following the
layer-by-layer growth, several $C_{\ell}(m)$ curves increases at the same time.
The system undergoes a wetting-layer and island-growth mode, or
Stranski-Krastanov growth mode. 

We also calculated the real-time roughness $R(m)$ of the film surface.  The
definition of $R(m)$ is the square root of the variance of the height of the
atoms homogeneously sampling on the surface.  The results are plotted in
Fig.~\ref{coverage&roughness} in thick curves in unit of the
thickness $d_L$ of a film layer.   It clearly shows that the roughness is
reduced by increasing temperature.  At 80K, $R(m)$ is about $2.5 d_L$ after 
deposition of 16 ML, because of the island growth; but at 600K, it is
ca.~$0.7d_L$, owing to the flat-film growth.  We have examined other systems,
such as Pt/Cu ($f=-7.91\%$) and Pd/Ni ($f=-9.51\%$). The results showed similar
trends of behavior: the coverage and the surface flatness are both improved by
increasing the substrate temperature. 

\begin{figure}[htbp]
\includegraphics[width=\figurewidth,angle=270]{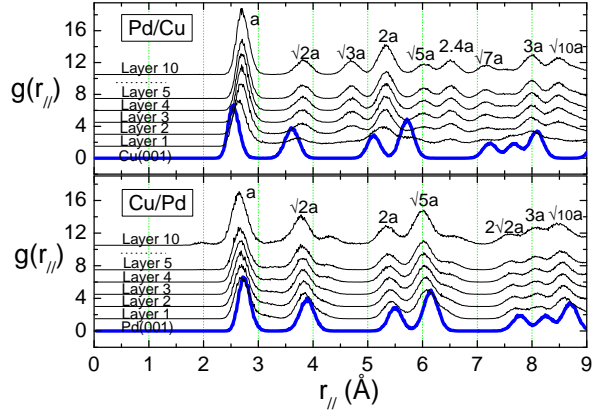}
\caption{$g_{\ell}(r_{\parallel})$ for Pd/Cu and Cu/Pd at 300K.  The number on
the left-hand side of a curve denotes the layer number $\ell$.  For the reason
of clarity, we have drawn the curves separately by shifting them vertically.
$g_{\ell}(r_{\parallel})$ for the two substrate surfaces, Cu(001) and Pd(001),
are plotted as references.  The positions of peaks in $g_{10}(r_{\parallel})$
are indicated, in turn, by  $a$, $\sqrt{2}a$, and so forth.} \label{2drdf}
\end{figure}

We then investigated the structure of thin film after deposition by analyzing
the two-dimensional radial distribution function (2D RDF) of each epilayer.
The 2D RDF is calculated by
$g_{\ell}(r_{\parallel})=\sigma_{\ell}(r_{\parallel})/\sigma_{\ell 0}$, where
$\sigma_{\ell}(r_{\parallel})$ is the surface density of adatom in the
$\ell$th epilayer at a radial distance $r_{\parallel}$ away from an adatom, and
$\sigma_{\ell 0}=N_{\ell}/L^2$ is the average surface density in the epilayer.  
In order to study the tensile-compressive asymmetry, $g_{\ell}(r_{\parallel})$
for the two counter-systems, Pd/Cu ($f=-7.20\%$) and Cu/Pd ($f=7.76\%$), at
300K were calculated. The results are illustrated in Fig.~\ref{2drdf}.

Focus firstly on the tensile case, Cu/Pd.  The referenced curve for the Pd
substrate surface shows the standard 2D RDF for a (001)-plane in which the peaks
appear subsequently at the positions $a$, ${\sqrt 2} a$, $2a$, ${\sqrt 5} a$,
\textit{etc.}, with $a$ being the nearest atomic distance.  We have verified
that the integration of these peaks with the appropriate weight $2\pi
r_{\parallel}\sigma_{00}$ reproduces, respectively, the numbers of the nearest,
the next nearest, the 3rd nearest neighbor atoms, and so forth.  We found that
the Cu epilayers display a $g_{\ell}(r_{\parallel})$ similar to the structure
of Pd surface.  It suggests a pseudomorphic growth of thin film along the
[001]-direction.  Since atomic spacing of Cu is smaller than Pd, the peaks in
the Cu epilayers move slightly toward left, as increasing the layer number, to
reduce the strain.  It approaches the bulk spacing $2.55$\AA\  of copper at
the 10th epilayer. 

\begin{figure}[htbp]
\includegraphics[width=\figurewidth,angle=270]{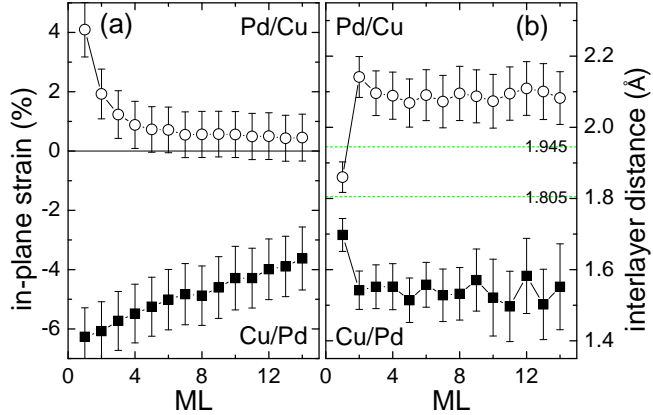}
\caption{(a) in-plane strain and (b) interlayer distance, during the film growth 
process} 
\label{in-plane-strain} 
\end{figure}

Focus secondly on the counter system, Pd/Cu.  The Pd epilayers now suffer from
compressive strain and release the strain by dislocation or alternating the
film to a close-packed structure.  Dramatic change in the
$g_{\ell}(r_{\parallel})$ structure was observed, compared to the previous
case.  The change includes the change of the peak height and the appearance of
some new peaks in reference of the 2D RDF of the Cu substrate.  Note 
that a new peak appears at ${\sqrt 3} a$ since the 2nd epilayer.  It strongly
suggests a hexagonal arrangement of atoms because the typical 2D RDF for a
hexagonal lattice shows a series of peaks at the positions $a$, ${\sqrt 3} a$,
$2a$, ${\sqrt 7} a$, and so on.  Therefore, the epilayers take a mixed
structure combining the (001)-  and the (111)-planes.  Recent experimental 
study~\cite{lu05} showed that the tensile overlayer for Cu/Pd(100) remains 
coherent up to about 9 ML, whereas for Pd/Cu(100), misfit dislocation appears 
just after few ML.  To make a  comparison with the experiments, 
we calculated in Fig.~\ref{in-plane-strain} the in-plane lattice strain 
(defined as $(a_{\rm bulk}-a_{\rm film})/a_{\rm bulk}$) and 
the interlayer distance of film during the growth process. 

We observed the striking asymmetric behavior in the relief of in-plane strain:
the compressive strain relaxes within just few ML, much more quickly than the
tensile strain does. This is in agreement with the experiments~\cite{lu05}.
For interlayer distance, a sudden increase appears for the Pd/Cu system while
the film grows more than one ML.  This is because Pd has larger lattice
constant than Cu; therefore, the distance between the Cu-Pd interface is
smaller than the Pd-Pd epilayer distance inside the film.  Similar argument can
also explain the sudden decrease of the interlayer distance at one ML for the
Cu/Pd system.  Moreover, the interlayer distance inside the Pd film is about
2.1\AA.  This value is larger than $1.945$\AA, the layer distance of a bulk
palladium along the [001] direction, but smaller than $2.246$\AA, the distance 
along the [111] direction. Since the in-plane strain relaxes quickly in
this case, this intermediate value reflects the fact of the formation of 
film layers, each of which takes a mixed planar (001)- and (111)-structure.
On the other hand, the interlayer distance
inside the Cu film is smaller than $1.805$\AA, half of the lattice constant of
copper. Since Cu film grows pseudomorphically on Pd(001) substrate, this small
value indicates the falling-down of an epilayer above another epilayer inside the film 
due to the unrelieved strain which has a in-plane lattice constant larger than the bulk one.
The interlayer distance is therefore reduced. Scientists~\cite{lu05} have observed 
exactly what we found here. 
Our results show good agreements with experiments in many aspects, more than qualitatively, 
although the deposition rate is fast.
This gives a direct support to our simulations  and the standard MD techniques 
can capture the main trend of behavior of this kind of systems and provide valuable 
information and insight into the problems. 

\begin{figure}[htbp]
\includegraphics[width=0.48\textwidth,angle=0]{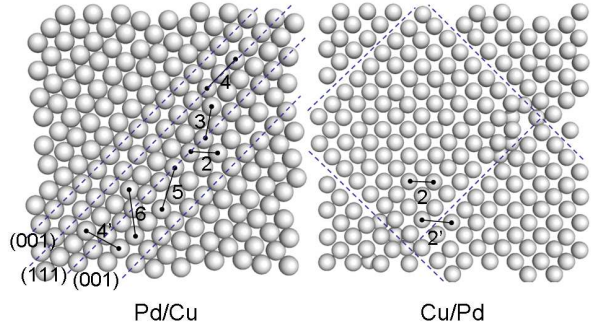}
\caption{Snapshots of the 10th epilayers for Pd/Cu and Cu/Pd at 300K.  Some
atomic distances which contribute the peaks in $g_{\ell}(r_{\parallel})$ in
Fig.~\ref{2drdf} have been plotted and numbered.  Dashed lines denote domain
boundaries.  } \label{shapshots} 
\end{figure}

In order to have a direct image of the deposited layer structure in mind,  we
present in Fig.~\ref{shapshots} the snapshots of the 10th epilayer for the two
systems.  
The Pd epilayer shows lattice displacement along the diagonal ([110]) direction
to reduce compressive strain, resulting in a series of structural changes 
between (001)- and (111)-planar structures, in form of strips.  
In the snapshot, some typical atomic distances,
which contribute the peaks in the RDF in Fig.~\ref{2drdf}, are plotted and
marked by the peak number.  For example, the 2nd and the 5th peaks in
$g_{\ell}(r_{\parallel})$ are derived from the atoms in the (001)-planar
strips, with the atomic arrangements similar to the links marked by 2 and 5,
respectively, in the picture; the 3rd peak comes from the arrangement in the
(111)-planar strips, similar to the link 3.  The links across the boundary of
the (001)- and (111)-planar strips contribute other peaks not appeared in the
standard RDFs, such as the peak at $r_{\parallel}=2.4a$ which is due to the
link 6.  The link 4' is across the strip boundary and has the length $1.93a$.
Nonetheless, lattice vibration at 300K smears the peak out, which becomes
indistinguishable with the 4th peak at position $2a$. 

The 10th epilayer for the Cu/Pd system shows much simpler structure, basically
following the structure of the (001)-plane.  It is an epitaxial growth.
However, the epilayer feels tensile strain due to the Pd substrate.  The strain
is relaxed by creating breaks in the diagonal directions, separating the film
layer into domains.  This kind of rectangular domains has been experimentally
observed~\cite{lu05,mitlin04}.  Moreover, we found that some atoms can sit on
the breaks with the atom height deviating of the layer, which results in the
small hump in the RDF curve near $r_{\parallel}=2.0$\AA.  For a link across a
domain boundary, like the link 2',  the length of the link is larger than the
link inside a domain, like link 2.  It is why small humps additionally appear
in the RDF in Fig.~\ref{2drdf}, for e.g., at $r_{\parallel}=4.3$\AA\, and at
other places.   

Paniago \textit{et al.}~\cite{paniago04} have studied the stress relaxation by
monitoring the RHEED pattern of the growth of Pd film on Cu(111) substrate.
They found a quick increase of atom spacing with layer number during
deposition, and the pseudomorphic growth was only up to the 2nd film layer.
Our Pd/Cu system shows consistent results.  Since our substrate surface is the
(001)-plane but not the (111)-plane as used in their experiment, one more
option exists for the film to relax the strain by transforming the structure
into the close-packed one.  We have verified other systems with smaller
compressive strain, for example, the Cu/Ni ($f=-2.49\%$) system. The film layer
turns to follow basically the structure of the nickel.  These findings
demonstrate the importance of lattice mismatch on thin film deposition in
pseudomorphic growth and in determination of the structure of epilayer.  
 
\begin{figure}[htbp]
\includegraphics[width=\figurewidth,angle=270]{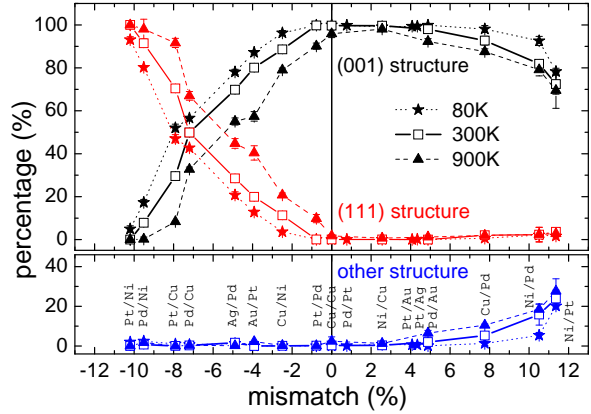}
\caption{Average percentages for the (001)-, the (111)-,  and the other
structures in an epilayer at 80K, 300K and 900K as a function of lattice
mismatch.} \label{percentage_misfit}
\end{figure}

Known that an epilayer can take a mixed structure while depositing onto a
(001)-plane, an important question to be answered is to quantify the
percentages of the (001)- and the (111)-planar structures in an epilayer and to
study  how they depend on the lattice mismatch.  We propose the following
method to achieve this goal.  The peaks at $r_{\parallel}={\sqrt 2}a$ and
${\sqrt 3}a$ in $g_{\ell}(r_{\parallel})$ can serve as an unambiguous
identifier to identify, respectively, the (001)- and the (111)-planar
structures, contributed from the next nearest neighboring atoms.  By
integrating the two peaks separately with weighting $2\pi
r_{\parallel}\sigma_{\ell 0}$, we calculated the mean numbers of the adatoms in
an epilayer with a distances ${\sqrt 2}a$ and ${\sqrt 3}a$, respectively, to an
adatom.  The percentages of the two structures were then computed by dividing
the two mean numbers separately by 4 and by 6, which are, in turn, the numbers of
the next nearest neighbors for a perfect (001)- and a perfect (111)-structure.
Subtracting these two percentages from 1 gives the percentage for the other
structure.  The results, averaged from the 5th to the 10th epilayer, which
excludes the possible interface alloying region, are shown in 
Fig.~\ref{percentage_misfit} as a function of lattice mismatch at three
temperatures 80K, 300K and 900K.

We observed an asymmetric curve with respect to $f=0$.  When $f$ is negative,
the percentage of the (001) structure decreases enormously with increasing
$|f|$; at the same moment, the percentage of the (111) structure increases.
Nearly no atom sits on a site other than the two structures.  The results shows
that for a large mismatch, such as the Pt/Ni system ($f=-10.2\%$), the film
layers tend to grow a close-packed structure to reduce the compressive strain,
instead of epitaxially following the square lattice of the substrate surface.
On the other hand, for a positive lattice mismatch, the percentage of the
(001)-structure also decreases with $f$, but more weakly compared to the
systems with negative mismatch.  In this case, the formation of the (111)-film
structure will not be helpful to reduce the tensile strain; thus, almost no
close-packed structure is formed.  The strain is relaxed by the creation of
breaks on the film layers.  The adatoms sitting on the breaks
(cf.~Fig.~\ref{shapshots}) contribute the percentage of the other structure.
This tensile-compressive asymmetry has been emphasized and discussed in recent
studies~\cite{lu05}. 

Concerning the effect of temperature, we found that the higher the substrate
temperature, the less the film will grow in a pseudomorphic way; in other
words, the percentage of the (001)-film structure decreases.  This effect is
more obvious in the regime of negative mismatch than positive one.  For
positive mismatches, the percentage of other structure also increases with
temperature.  We know that the (111) plane has the lowest surface energy for
fcc metals.  Increasing temperature enhances the migration of an adatom to a
more thermodynamically stable site, the site on the (111) plane.  A negative
misfit will promote this migration to form a close-packed structure  to reduce
the mechanically compressive strain, whereas a positive misfit will hinder it
because it increases the tensile strain which is mechanically unfavorable.  The
interplay between the thermodynamics and the mechanics of the materials
determines the film structure. 

For the implication to experiments, this study suggests that it is easier to 
grow pseudomorphic film on (001) surface of these transition metals
in the positive region of lattice mismatch than in the negative. 
We also emphasize that the method proposed in this study to quantify the 
structural composition can be adapted to experiments because the 2D RDF of
a film layer can be experimentally obtained by surface diffraction techniques.
Thus the structural composition versus lattice mismatch can be calculated to
understand this strong asymmetry predicted by our simulations. Concerning the 
limitation of the standard MD techniques, it is definitely worth in the future
to investigate the effect of the neglected interlayer diffusion processes on 
this problem using more sophisticated simulation techniques. 

In summary, we have discussed the effect of lattice mismatch on the structure
of epilayers on the (001) substrate surface  among six transition metals.  The
layer coverage, the surface roughness, and the temperature effect have also
been studied.  We presented the first calculation of the composition percentage
of film structure for positive and negative lattice mismatches.  The results
revealed a strong asymmetric behavior between the tensile and the compressive
cases.  The epilayer releases compressive strain by the formation of the
close-packed planar structure, but release tensile strain by the creation of
breaks and domains on a film layer.    

This work is supported by the National Science Council, the Republic of China,
under the contract No.~NSC 95-2112-M-007-025-MY2.  Computing resources are
supported by the National Center for High-performance Computing.

\end{document}